\documentclass[a4paper,12pt]{article}
\usepackage{epsfig}
\newcommand {\kgdt}  {9.31}
\newcommand {\kgdc}  {5.03}
\newcommand {\kgdca} {3.80}
\newcommand {\kgdcb} {0.63}
\newcommand {\kgdcc} {0.60}

\newcommand {\kgdeff}{4.53}
\newcommand {\neff}{0.51}
\begin{document}
\begin{titlepage}

\begin{flushright}
       LYCEN 2001-38  \\
       \today
\end{flushright}

\vfill

{\bf\LARGE
\begin{center}
           First Results of the EDELWEISS
           WIMP Search using a 320~g
           Heat-and-Ionization Ge Detector
\end{center}}

\vfill

\begin{center}
{\large The EDELWEISS Collaboration:} \\
  A.~Benoit$^{1}$,
  L.~Berg\'e$^{2}$,
  A.~Broniatowski$^{2}$,
  B.~Chambon$^{3}$,
  M.~Chapellier$^{4}$,
  G.~Chardin$^{5}$,
  P. Charvin$^{5,6}$,
  M.~De~J\'esus$^{3}$,
  P. Di~Stefano$^{5}$,
  D.~Drain$^{3}$,
  L.~Dumoulin$^{2}$,
  J.~Gascon$^{3}$,
  G.~Gerbier$^{5}$,
  C.~Goldbach$^{7}$,
  M.~Goyot$^{3}$,
  M.~Gros$^{5}$,
  J.P.~Hadjout$^{3}$,
  A.~Juillard$^{2,5}$,
  A.~de~Lesquen$^{5}$,
  M.~Loidl$^{5}$,
  J.~Mallet$^{5}$,
  S.~Marnieros$^{2}$,
  O.~Martineau$^{3}$,
  N.~Mirabolfathi$^{2}$,
  L.~Mosca$^{5}$,
  L.~Miramonti$^{5}$,
  X.-F.~Navick$^{5}$,
  G.~Nollez$^{7}$,
  P.~Pari$^{4}$,
  M.~Stern$^{3}$,
  L.~Vagneron$^{3}$
\end{center}

{\scriptsize\noindent
$^{1}$Centre de Recherche sur les Tr\`es Basses Temp\'eratures,
      SPM-CNRS, BP 166, 38042 Grenoble, France\\
$^{2}$Centre de Spectroscopie Nucl\'eaire et de Spectroscopie de Masse,
      IN2P3-CNRS, Universit\'e Paris XI,
      bat 108, 91405 Orsay, France\\
$^{3}$Institut de Physique Nucl\'eaire de Lyon-UCBL, IN2P3-CNRS,
      4 rue Enrico Fermi, 69622 Villeurbanne Cedex, France\\
$^{4}$CEA, Centre d'\'Etudes Nucl\'eaires de Saclay,
      DSM/DRECAM, 91191 Gif-sur-Yvette Cedex, France\\
$^{5}$CEA, Centre d'\'Etudes Nucl\'eaires de Saclay,
      DSM/DAPNIA, 91191 Gif-sur-Yvette Cedex, France\\
$^{6}$Laboratoire Souterrain de Modane, CEA-CNRS, 90 rue Polset,
      73500 Modane, France\\
$^{7}$Institut d'Astrophysique de Paris, INSU-CNRS,
      98 bis Bd Arago, 75014 Paris, France
}

\vfill

\begin{center}{\large\bf Abstract}\end{center}

The EDELWEISS collaboration has performed a direct search for WIMP dark
matter using a 320~g heat-and-ionization cryogenic Ge detector operated
in a low-background environment in the Laboratoire Souterrain de Modane.
No nuclear recoils are observed
in the fiducial volume
in the 30-200 keV energy range
during an effective exposure of \kgdeff\ kg$\cdot$days.
Limits for the cross-section for the spin-independent interaction
of WIMPs and nucleons are set in the framework of the Minimal
Supersymmetric Standard Model (MSSM).
The central value of the signal reported by the experiment
DAMA is excluded at 90\% CL.

\vfill


\end{titlepage}


\noindent{\large\bf Introduction}

A general picture of matter and energy in the Universe
is now emerging (see e.g. Ref.~\cite{bib-review}
for a review),
suggesting that our Galaxy could be immersed
in a halo of Dark Matter
made of Weakly Interacting Massive Particles (WIMPs).
The collision of a WIMP with an atomic nucleus would produce a nuclear
recoil with a kinetic energy of the order of ten keV~\cite{bib-sandl}. 
In the event that WIMPs are the neutralinos of the Minimal Supersymmetric
extension of the Standard Model (MSSM), interaction rates
per kilogram of matter
would vary between 1 event per day to one per decade,
depending on model parameters~\cite{bib-mssm}.

Experimental searches for these recoils in germanium
ionization detectors~\cite{bib-ge}
and NaI scintillators~\cite{bib-nai}
have yielded upper limits on their rate per kilogram of detector material,
which are interpreted in the framework of the MSSM in terms of limits
on the WIMP-nucleon interaction cross-section.
These searches are limited by the interaction rate due
to natural radioactivity, which is at best limited to
approximately 1 count/kg/day in the low
energy range where recoils are expected.

Recently, the DAMA experiment has reported an annual modulation of
the low-energy rate recorded in their $~100$ kg NaI detector array
over a period of four years~\cite{bib-dama}.
This was attributed~\cite{bib-dama} to the modulation of
the WIMP flux impinging on the detector due to the
Earth rotation around the Sun corresponding to a
WIMP mass of 52$\pm^{10}_{8}$ GeV/c$^2$ and a WIMP-nucleon interaction
cross-section of (7.2$\pm^{0.4}_{0.9}$) $\times$10$^{-6}$pb.
In contrast, the CDMS collaboration~\cite{bib-cdms}
observed no excess of nuclear recoils above the rate
expected from the scattering of cosmic-ray induced neutrons
after accumulating an exposure of 
10.6 kg$\cdot$days in the fiducial volume of their
heat-and-ionization cryogenic germanium detectors.
The two experimental results are not compatible if
one applies the standard procedure to scale rates
in different detectors described in ref.~\cite{bib-sandl}.

More data are needed to resolve definitely this discrepancy.
The most exciting developments here are to be expected from
the rapidly evolving domain of heat-and-ionization
(or heat-and-scintillation~\cite{bib-cresst}) cryogenic
detector technology, which provides excellent event-by-event
rejection of the dominating $\gamma$-ray background.
The EDELWEISS collaboration has recently commissioned
a massive (320~g) heat-and-ionization Ge detector~\cite{bib-navick}.
We report here on the first physics results obtained with
this detector in a low-background environment in the underground
site of the Laboratoire Souterrain de Modane (LSM).
In this deep-underground experiment,
the cosmic-ray induced neutron background
that limited the recent CDMS results
(one nuclear scattering per kg$\cdot$day above 10 keV recoil
energy~\cite{bib-cdms})
should be reduced by orders of magnitude.
The CDMS and EDELWEISS detectors differ by their mass,
geometry and electrode implantation scheme,
and a comparison of their performance will benefit
the development of this novel technology.
The results presented in this letter represent a significant
improvement relative to our previous
results~\cite{bib-distefano,bib-benoit},
obtained with a 70~g detector and with higher
background levels.

\noindent{\large\bf Experimental Setup}

The experimental site is the Laboratoire Souterrain de Modane
in the Fr\'ejus Tunnel under the French-Italian Alps.
The 1780~m rock overburden (4800~m water equivalent) results in
a muon flux of about 4 m$^{-2}$day$^{-1}$ in the experimental
hall and the flux of neutrons in the 2-10 MeV range is
\mbox{4$\pm$1 $\times$10$^{-6}$ cm$^{-2}$s$^{-1}$~\cite{bib-chazal}}. 


The detector is mounted in a dilution cryostat
shielded from the radioactive environment
by 10~cm of copper and 15~cm of lead~\cite{bib-cryostat}.
Pure nitrogen gas is circulated around the cryostat in order
to reduce radon accumulation.
The radioactivity of all material in the close vicinity of the
detectors was measured using a dedicated low-background
germanium $\gamma$-ray detector, also at the LSM.
All electronic components were moved away from the
detector and hidden behind a 7~cm thick archeological lead shield.
The entire setup is surrounded by a 30~cm thick
paraffin shielding against neutrons.
According to Monte Carlo simulations of the various
shields based on the measured neutron flux in the
experimental hall,
the rate of neutron scattering events producing
nuclear recoils above 30 keV is expected to be of
the order of 0.03 per kg and per day.


The detector~\cite{bib-navick} is a cylindrical Ge single crystal
with a diameter of 70~mm and a thickness of 20~mm.
The edges have been beveled at an angle of 45$^o$.
The plane surfaces and chamfers have been metalized for
ionization measurement.
The electrodes are made of 100~nm Al layers sputtered on the
surfaces after etching.
The top electrode is divided in a central part and a guard ring,
electrically decoupled for radial localization of the
charge deposition.
During data taking, a voltage $V_o =$ 6.37 V was applied to
the top electrodes.
The electrodes were regularly shorted in order to
prevent charge accumulation due to trapping of carriers
in the detector volume.

The cross-talk between the centre and guard ring electrode
signal is approximately 10\%.
This does not affect the ionization energy resolution since
the cross-talk amplitude is a fixed fraction of the
signal amplitude on the other electrode.
Its shape is constant in time and this effect is
easily taken into account by the simultaneous analysis of
the signals recorded on both electrodes.

The thermal sensor consists of a Neutron Transmutation Doped
germanium crystal (NTD) of 4 mm$^3$ glued to the beveled
part of the surface of the detector.
The residual radioactivity of the activated NTD sensor should
thus be mostly contained in the region covered by the guard
electrode.
The resistance of the DC-polarized sensor was approximately
3 M$\Omega$ for a base temperature of 27 mK,
stabilized to within $\pm$10~$\mu$K.

The signals from all channels are sent to digitizers triggered
by either of the two ionization channels.
To generate the trigger, the ionization signals are
processed by shaping amplifiers and sent to
two discriminators.
The trigger is the ``or'' of the output of the
two discriminators.
The rise time of the heat signal is of the order
of 10~ms, much slower than the $\mu$s-fast
ionization channel, and was not used for triggering.

\noindent{\large\bf Detector Calibration}

The heat and ionization responses to $\gamma$ rays were
calibrated using $^{57}$Co sources.
Over the entire data-taking period,
the baseline resolutions on the centre and guard ring
ionization signals are better than 2.0 keV FWHM,
and it varied between 1.9 and 3.5 keV FWHM for the heat
signal.
The corresponding resolutions measured at 122 keV
for the centre, guard ring and heat signals are
approximately
3, 2 and 3.5 keV FWHM, respectively.

The ionization resolution was limited by
microphonic noise at frequencies varying with time.
This also constrained the trigger level of
the data acquisition, which was set on
the relatively faster ionization signals.
The individual trigger levels on the two ionization
channels were repeatedly measured during the data
taking period using the low-energy part of the Compton
plateau recorded with a $^{60}$Co source.
The low-energy edge of this plateau corresponds to
the decrease of efficiency due to the trigger threshold.
Its shape is well described by an error function
corresponding to a 50\% efficiency
at 5.7$\pm$0.3 keV ionization and reaching full
efficiency at approximately 8 keV ionization.

\begin{figure}[tbp]
\epsfig{file=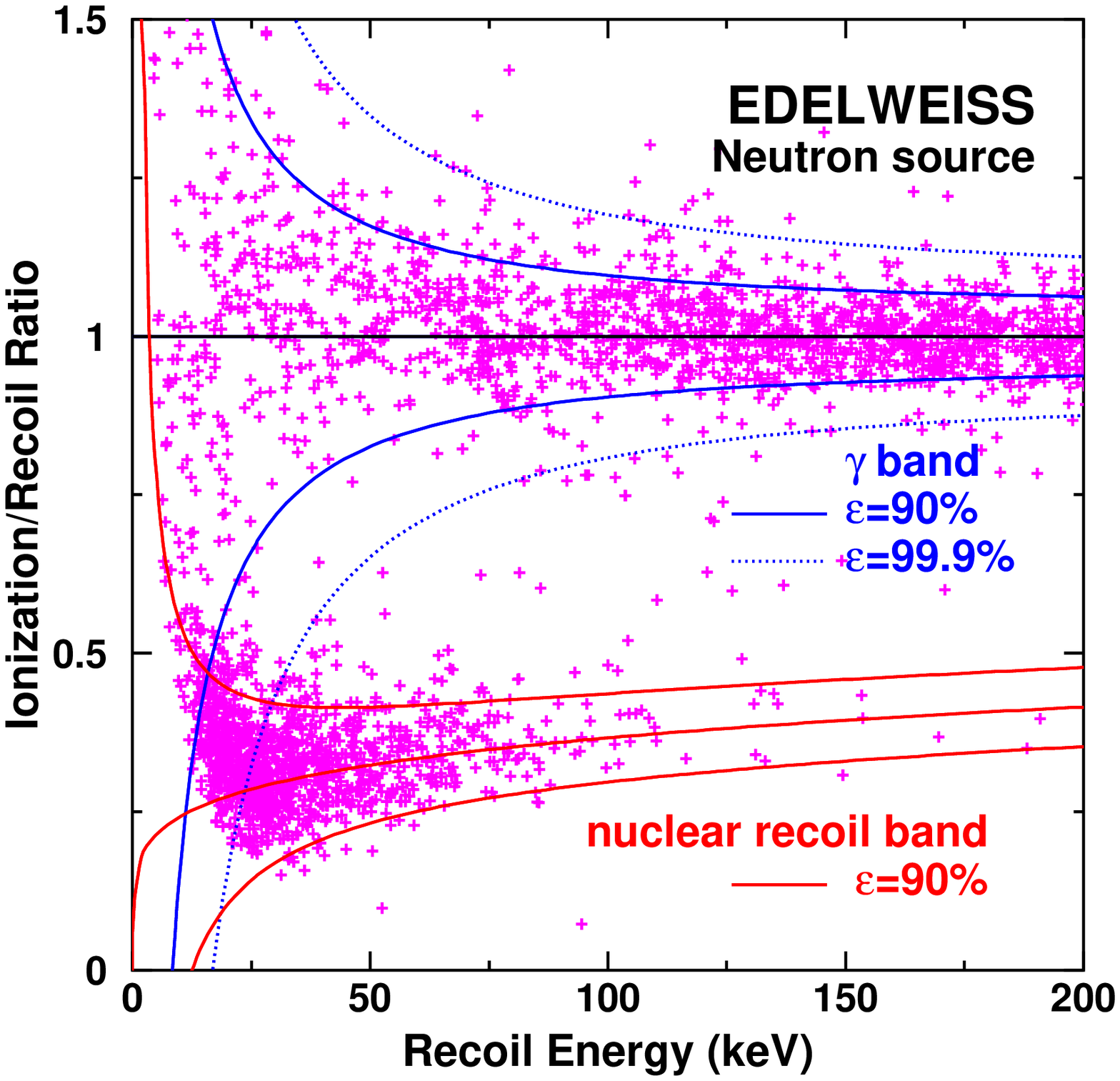
,height=17cm,bbllx=1cm,bblly=5cm,bburx=23cm,bbury=27cm}
\caption[]{Distribution of the quenching factor
(ratio of the ionization signal to recoil energy)
as a function of the recoil energy
from the data collected in the centre fiducial volume during
the neutron calibration run of a 320~g EDELWEISS detector
using a $^{252}$Cf source.
The full lines correspond to the $\pm$1.645$\sigma$ bands
(90\% efficiency) for photons and for nuclear recoils
and the dotted lines the $\pm$3.29$\sigma$ band
(99.9\% efficiency) for photons.}
\label{fig-neutron} 
\end{figure}

The calibration of the response to photons and
nuclear recoils was performed using $^{57}$Co
and $^{60}$Co $\gamma$-ray sources and a
$^{252}$Cf neutron source.
The variable used to discriminate electron and
nuclear recoils is the ratio of the ionization
signal to the recoil energy.
To obtain the recoil energy $E_R$, the
heat signal is corrected for the Joule
heating proportional to the charge signal
amplitude~\cite{bib-luke}.
For this, the ionization and heat signals
calibrated using $\gamma$-ray sources
(the electron-equivalent energies $E_I$ and $E_H$,
respectively) are combined event-by-event
to get the true recoil energy $E_R$
\begin{eqnarray}
E_R & = & (1+\frac{V_o}{V_{pair}})E_H
           - \frac{V_o}{V_{pair}} E_I \nonumber
\end{eqnarray}
where $V_{pair}=3~V$ is the electron-hole pair creation
potential in Ge.
By construction, the ratio of the ionization energy to the recoil
energy, $Q = E_I / E_R$, is equal to 1 for energy deposits coming
from $\gamma$-rays.
For neutrons, $Q$ is a function of $E_R$ determined
using the data recorded with a $^{252}$Cf source
shown in Fig.~\ref{fig-neutron}.
The average value of $Q(E_R)$ is well described
by $Q=0.16(E_R)^{0.18}$, with $E_R$ in keV,
a parameterization
similar to that obtained elsewhere~\cite{bib-qn}.
From the dispersion of these data is obtained the
{\em nuclear recoil band}, defined as the region in
the ($Q$,$E_R$) plane where 90\% of the nuclear
recoils are expected.
Since neutrons have a significant probability for multiple
scattering inside the detector volume,
the $Q(E_R)$ distribution for an actual WIMP signal
would differ slightly from that measured with the
$^{252}$Cf source.
However, Monte Carlo simulations indicate that multiple
scattering shifts down the average measured Q values for
neutrons by approximately 0.01 units relative to a WIMP
signal.
This value is small compared to the width of the adopted
nuclear recoil band shown in Fig.~\ref{fig-neutron}
and has been neglected.
Furthermore, the small decrease of the WIMP detection
efficiency would be largely compensated by the narrowing
of the $Q$ distribution at a given recoil energy due to the
absence of multiple scattering.


An important feature of the detector is the ability to
use the guard electrode to tag interactions
occurring near the perimeter of the detector.
This part of the surface is the most exposed to elements
of the detector environment that are known to have perceptible
levels of radioactivity (such as the NTD and Cu-Be support
springs)
and its surface-to-volume ratio makes it more
susceptible to surface contaminants.
Interactions in this region can also suffer from
electric field inhomogeneities leading to incomplete charge
collection,
and thus mimic the ionization deficit expected for
nuclear recoils.
For this reason, the fiducial volume of the detector
is chosen to correspond to events for which more
than 75\% of the charge is collected in the centre
electrode.

The fiducial volume fraction $f_V$ corresponding to this
selection was measured using nuclear recoil events
recorded with the $^{252}$Cf source,
as neutron interactions are expected to be more evenly
spread throughout the detector than low-energy
$\gamma$-ray interactions.
A clean sample of neutron events is obtained
from the $^{252}$Cf data by requiring Q$<0.5$
and a recoil energy between 30 and 200 keV.
These energies correspond to an ionization well above the 
trigger threshold, and the loose requirement on $Q$
makes the selection insensitive to difference in
resolution between the two electrode signals.
In this sample where the selection efficiency does not
depend on the relative strength of the signal on the two
electrodes,
the fiducial volume cut selects
$f_{event}$ $=$ 53$\pm$2\% of the events.
Some sharing of the charge between the two electrodes
is expected to arise because of multiple scattering events
occurring both inside and outside the fiducial volume
and of interactions occurring close to the boundary between
the two electrodes.
Multiple scattering must thus be taken into account
in the derivation of the volume fraction from the measured
fraction of events passing the fiducial cut.
To evaluate this correction, the response of the
detector, the cryostat and the shielding to
neutrons from the $^{252}$Cf source was
simulated using GEANT~\cite{bib-geant}.
According to the simulations, the volume fraction $f_V$
corresponding to the fiducial selection is 54$\pm$2\%.
This number differs from the event fraction
$f_{event}$ by
only 1\% because
the fiducial selection partially compensates
the loss of ``pure centre'' events due to
multiple scattering
by allowing that up to 25\% of the charge be
collected on the guard ring electrode.

Simple electrostatic simulations of the bending of
the field lines inside the detector can reproduce
the volume fractions to within 5\%.
This discrepancy is taken as a systematic error on
the measured fraction, which is then
$f_V$ $=$ 54 $\pm$2 (stat.) $\pm$5 (syst.) \%.

\noindent{\large\bf Results and Discussion}

The low-background data were accumulated
over two consecutive months.
Over these two months, no physics runs have been excluded from
the data sample.
In two series of runs, an increase of the microphonic noise level
on the centre electrode channel lead to unacceptably high
trigger rates.
These were brought under control by attenuating the centre
electrode signal by a calibrated amount, increasing the
50\% efficiency level for these runs from 5.7 to 9 and 11 keV.
Using the value of $f_V$ obtained in the previous section,
the exposure recorded with 5.7, 9 and 11 keV ionization thresholds
are \kgdca, \kgdcb\ and \kgdcc\ kg$\cdot$day,
respectively, for a total of \kgdc\ kg$\cdot$day (fiducial volume).
The principal source of down-time were the regular interruptions
for calibrations with $\gamma$ sources and cryostat operations.

The data taking was interrupted by a series of power cuts
that were followed by a significant deterioration
of charge collection, as attested by the rate of events
below the nuclear recoil band.
No such events were observed between 30 and 200 keV recoil
energy before the incident
(corresponding to a rate inferior to 0.5 /kg/day at
90\% confidence level),
while the observed rate is 1.8$\pm$0.6 /kg/day afterwards.
Work is in progress to understand the exact cause
of this deterioration and to restore the original charge
collection properties of the detector.

\begin{figure}[tbp]
\epsfig{file=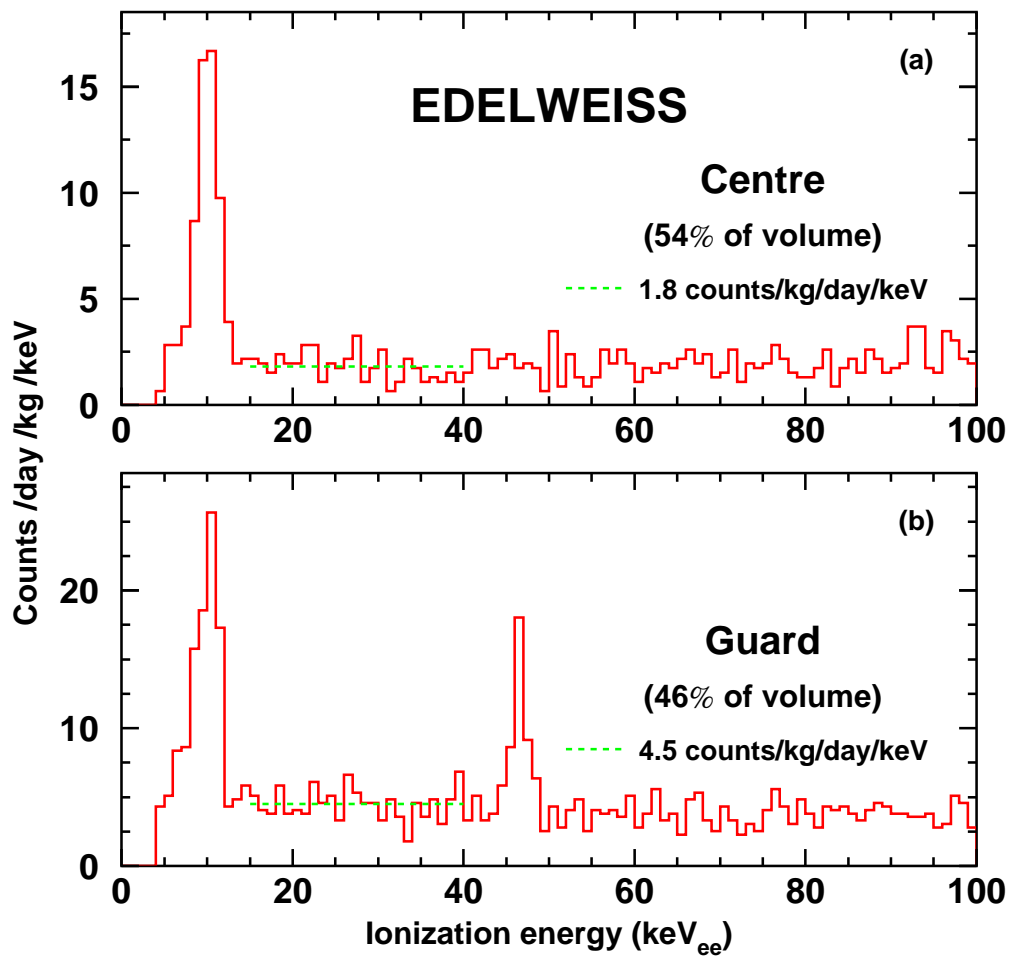
,height=17cm,bbllx=1cm,bblly=5cm,bburx=23cm,bbury=27cm}
\caption[]{Ionization pulse height spectra
recorded in the \kgdt\ kg$\cdot$day exposure of the 320~g
EDELWEISS detector: (a) spectrum recorded in the
centre fiducial volume; (b) spectrum recorded
in the rest of the detector.}
\label{fig-ion} 
\end{figure}

Fig.~\ref{fig-ion}a shows the distribution of
the total (centre+guard ring) ionization energy
recorded for events passing the fiducial volume cut,
before applying the nuclear recoil selection.
Fig.~\ref{fig-ion}b shows the corresponding
distribution for events rejected by the fiducial volume cut.
The 46.5 keV $^{210}$Pb line provides
a convincing illustration of the ability of the
fiducial volume selection to reject localized
sources of background radioactivity.
Indeed, its yield in the rejected sample is
3.4 $\pm$ 0.5 counts/day
while it is less than 0.3 counts/day in the fiducial
volume (at 90\% CL),
proving that the source of this contamination is located towards
the outside perimeter of the detector.

The 10.4 keV line observed in both spectra and
corresponding to the Ga K-shell energy originates
from the cosmogenic activation of the detector
leading to the creation of $^{68}$Ge with a half-life
of T$_{1/2}$=271 days~\cite{bib-nuclide}.
No significant variation of its intensity is observed
as a function of time, indicating that the contribution
from the decay of the $^{71}$Ge nucleus (T$_{1/2}$=11.2 days)
created by thermal
neutron capture during the exposure to the $^{252}$Cf source
is small.
In both cases, these decays should be evenly distributed
in the detector volume and thus provide an alternative tool
for the measurement of the fiducial volume fraction.
This method is statistically less precise, and the effect
of the finite trigger threshold must be taken into account.
Nevertheless, the value of $f_V$ obtained with this method
(50 $\pm$ 4~\%, where only the statistical error is quoted)
is compatible with the neutron data result.

Another illustration of the usefulness of the fiducial
volume selection is the reduction 
in the overall count rates between 15 and 40 keV ionization energy
observed in fig.~\ref{fig-ion},
from 4.5$\pm$0.2 to 1.8$\pm$0.1 counts/kg/day/keV.

\begin{figure}[tbp]
\epsfig{file=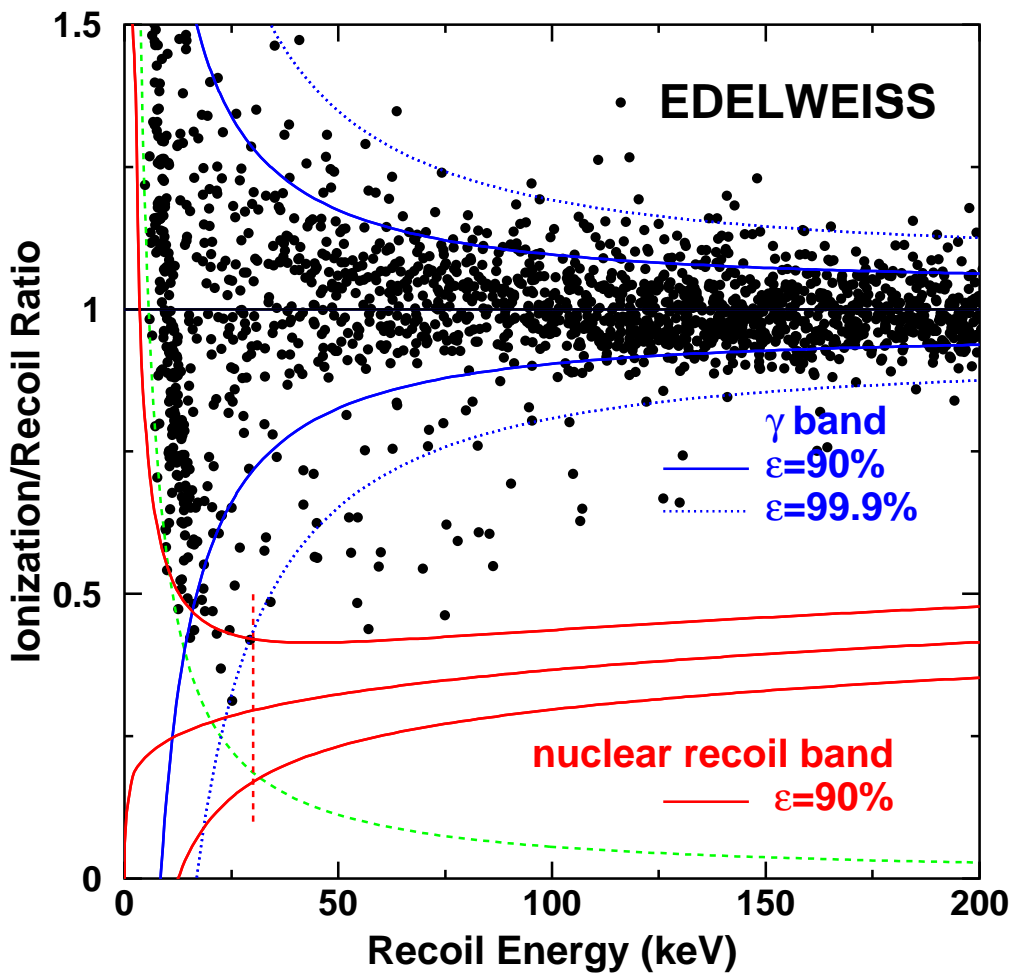
,height=17cm,bbllx=1cm,bblly=5cm,bburx=23cm,bbury=27cm}
\caption[]{Distribution of the quenching factor
(ratio of the ionization signal to the recoil energy)
 as a function of the recoil energy
from the data collected in the centre fiducial volume
of the 320~g EDELWEISS detector.
The exposure of the fiducial volume corresponds to \kgdc\ kg$\cdot$day.
Also plotted are the $\pm$1.645$\sigma$ bands (90\% efficiency) for
photons and for nuclear recoils.
The 99.9\% efficiency region for photons is also shown (dotted line).
The hyperbolic dashed curve corresponds to 5.7 keV ionization energy
and the vertical dashed line to 30 keV recoil energy.}
\label{fig-data} 
\end{figure}

Fig.~\ref{fig-data} shows the distribution of $Q$ versus $E_R$
for the entire data set.
Most events are within the 99.9\% efficiency photon band.
A few events lie between this region and the nuclear recoil
band.
They are interpreted as surface events with reduced charge
collection~\cite{bib-benoit}.
The present results are significantly better than those
obtained previously~\cite{bib-distefano,bib-benoit}
with smaller detectors.
Part of the improvements comes from the decrease of radioactive
backgrounds in the close vicinity of the detector,
as well as the mass of the detector (320~g, the largest mass
achieved so far for heat-and-ionization cryogenic Ge detectors)
which allows the definition of a large fiducial volume
with a relatively uniform electrostatic field
surrounded by a thick protective guard region.
The sputtered Al electrodes which equip the present detector
also appear to have reduced the charge collection problems
for surface events that severely limited the previous
detector performances~\cite{bib-distefano}.

The limit on the WIMP rate is taken from the total
number of counts in the nuclear recoil band
for recoil energies between 30 and 200 keV.
The lower limit corresponds to the recoil energy
for which the efficiency is close to 90\%
and excludes the region where the $\gamma$-rays rejection
is worse than 99.9\%,
and in particular the low-$Q$ tail of the 10.4 keV line.
The experimental limit on the rate between 30 and 200 keV
recoil energy is less than \neff\ counts per kg$\cdot$day
at 90\% CL.

\begin{figure}[tbp]
\epsfig{file=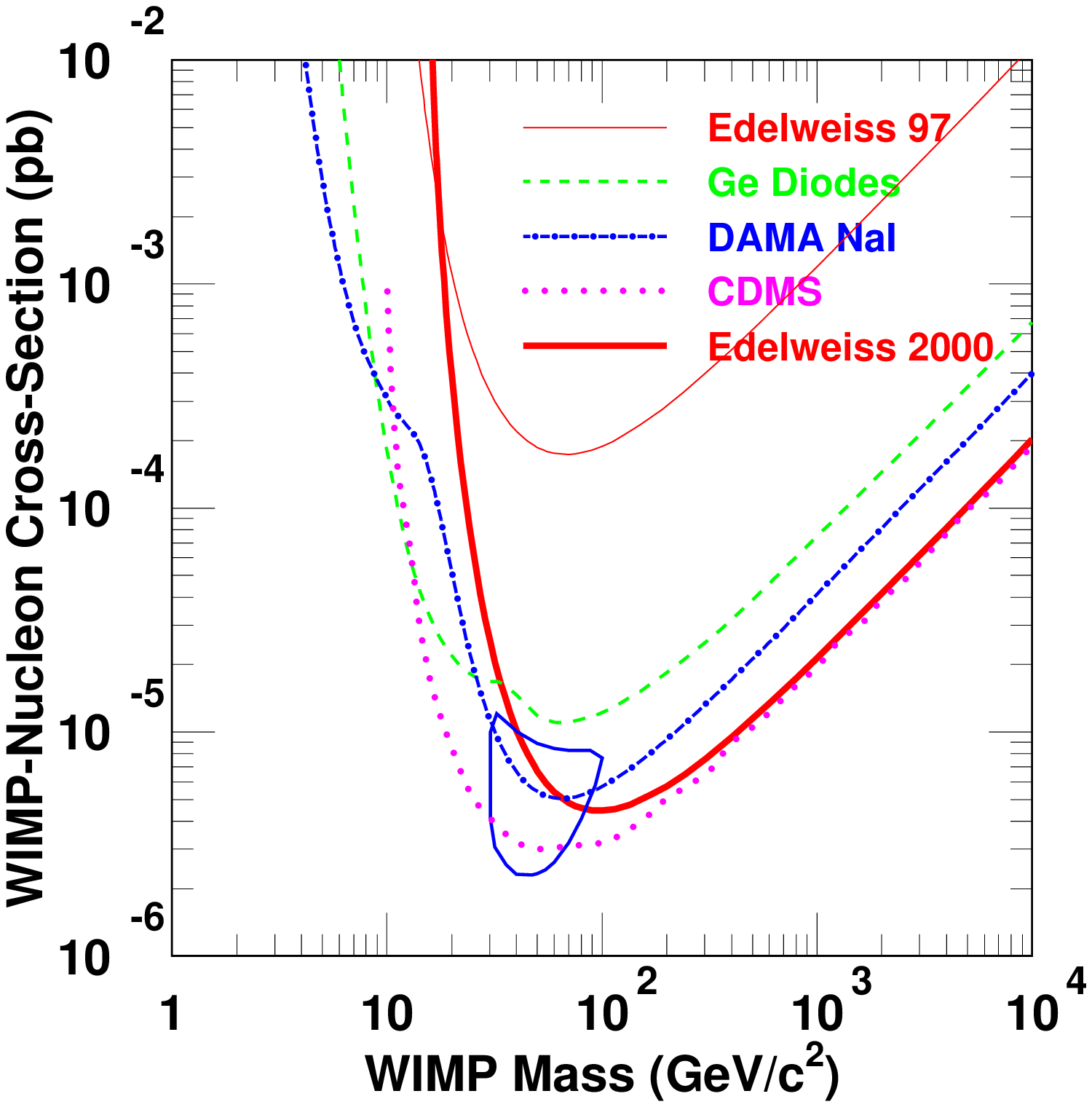
,height=17cm,bbllx=0cm,bblly=0cm,bburx=20cm,bbury=20cm}
\caption[]{Spin-independent exclusion limit (dark solid curve) obtained
in this work.
Thin solid curve: previous EDELWEISS results~\protect\cite{bib-distefano}.
Dashed curve: combined Ge diode limit~\protect\cite{bib-ge}.
Dash-dotted curve: DAMA NaI limit using pulse-shape
discrimination~\protect\cite{bib-nai}.
Dotted curve: CDMS limit with statistical subtraction
of the neutron background~\protect\cite{bib-cdms}.
Closed contour: allowed region at 3$\sigma$ CL
for a WIMP r.m.s. velocity of 270 km/s from the DAMA
annual modulation data~\protect\cite{bib-dama}.}
\label{fig-exclus} 
\end{figure}

These results are interpreted in terms of an upper
limit at 90\%~CL on the WIMP-nucleon scattering cross-section
using the prescriptions of Ref.~\cite{bib-sandl}.
The limits are shown in Fig.~\ref{fig-exclus}.
In the calculation of the WIMP flux, a galactic halo WIMP
density of 0.3 $GeV/cm^3$ is assumed, together with
an r.m.s. velocity of 270 km/s,
an escape velocity of 650 km/s and
a relative Earth-halo velocity of 230 km/s.
The interaction rates are calculated using cross-sections
scaled by the square of the target mass number
and the Helm parameterization of the form
factor~\cite{bib-sandl,bib-helm}.
The expected number of WIMPs as a function of
their mass and scattering cross-section takes into
account the experimental efficiency for nuclear
recoils as a function of the recoil energy 
and uses the ionization threshold values relevant
to each data sample.

To ensure the stability of this result within the
systematic uncertainties of the measurement,
the analysis has been repeated using an increased fiducial
volume.
It corresponds to the selection of all events where the
signal on the centre electrode is larger than that on the guard.
The fiducial volume evaluated from the neutron source data
is 63$\pm$2\%,
corresponding to a 17\% increase of the fiducial volume.
No nuclear recoil candidates are observed in the
increased-acceptance sample.
This increase in acceptance is larger than the variations 
corresponding to the uncertainties on the trigger threshold
and on the measurement of $f_V$.
A further increase in efficiency can be achieved by increasing
the width of the nuclear recoil band to 95\% efficiency
with still no events entering the band. 
It is thus believed that the limits shown in Fig.~\ref{fig-exclus}
are conservative.

For WIMP masses above 30 GeV/c$^2$,
the present limits are better than those obtained
by Ge diode experiments without heat measurement~\cite{bib-ge}.
Although the effective exposure in this experiment
is approximately half that accumulated by the
CDMS collaboration~\cite{bib-cdms}, 
the limits obtained for WIMP masses above 200 GeV/c$^2$
are very similar, as seen in  Fig.~\ref{fig-exclus}.
This is due to the absence of any event in the
EDELWEISS acceptance while the CMDS results relies
on a statistical subtraction of its neutron
background.
At lower WIMP masses, the present results suffer from
the relatively poorer energy resolution and high
ionization trigger level.
This should be solved with planned improvements in
the wiring of the detector to reduce microphonic noise
and adjustments of the NTD sensor excitation and readout.

Based on the usual assumptions for the
comparison of direct WIMP searches described in
Ref.~\cite{bib-sandl} such as target mass scaling,
and using the same standard galactic halo
model~\cite{bib-dama,bib-cdms}, 
the EDELWEISS results~\footnote{The EDELWEISS 90\% CL
cross-section limit for a WIMP mass of 52 GeV/c$^2$
is $\sigma_n$~=~6.3~$\times$10$^{-6}$pb.}
exclude at more than 90\% CL
the central value for the WIMP signal reported
by the annual modulation measurement of the
DAMA collaboration~\cite{bib-dama}
(WIMP mass $M_W$ = 52 GeV/c$^2$ and interaction
cross-section $\sigma_n$ = 7.2 $\times$10$^{-6}$pb).
It does not exclude at 90\% CL the other central
value obtained by DAMA when the annual modulation data
is combined with their own exclusion data based on
pulse shape discrimination
($M_W$ = 44 GeV/c$^2$ and
$\sigma_n$ = 5.4 $\times$10$^{-6}$pb).

\noindent{\large\bf Conclusion}

The EDELWEISS collaboration has searched for nuclear recoils
due to the scattering of WIMP dark matter using a 320~g
heat-and-ionization Ge detector operated in a low-background
environment in the Laboratoire Souterrain de Modane.
After an effective exposure of \kgdeff\ kg$\cdot$day,
the rate of Ge recoils
with kinetic energies between 30 and 200 keV
is measured to be less than \neff\  per kg$\cdot$day
at 90\% CL.
This is the most stringent limit based on the
observation of zero event and not relying on
any statistical background subtraction.
Within the usual assumption for the
comparison of direct WIMP searches~\cite{bib-sandl}
and using the same standard galactic halo
model as in Ref.~\cite{bib-dama,bib-cdms}, 
the EDELWEISS results exclude at more than 90\% CL
a 52 GeV/c$^2$ WIMP with an interaction cross-section
of 7.2 $\times$10$^{-6}$~pb.
With a four-fold increase of exposure time or with
some improvements in the detector resolution,
the sensitivity of the present detector should
be able to test the whole parameter space of the DAMA
candidate, without requiring a statistical
subtraction of nuclear recoils due to neutron
scattering interactions.
Later this year it is planned to operate 
at the LSM this detector
together with two other similar 320~g detectors
presently under construction.

\section*{Acknowledgements}
The help of the technical staff of the Laboratoire Souterrain
de Modane and the participating laboratories is gratefully acknowledged.
This work has been partially funded by the EEC Network program under
contract ERBFMRXCT980167.

\newpage

\begin{thebibliography}{99}
%
\bibitem{bib-review}
  M.S. Turner, Phys. Scripta {\bf T85} (2000) 210; \\
  L. Bergstr\"{o}m, Rep. Prog. Phys. {\bf 63} (2000) 793.
\bibitem{bib-sandl}
J.D.~Lewin and P.F.~Smith, Astropart. Phys. {\bf 6} (1996) 87.
\bibitem{bib-mssm}
  G. Jungman, M. Kamionkovski and K. Griest,
  Phys. Rep. {\bf 267} (1996) 195.
\bibitem{bib-ge}
  D. Reusser {\em et al.}, Phys. Lett. B {\bf 255} (1991) 143; \\
  L. Baudis {\em et al.}, Phys. Rev. D {\bf 59} (1998) 022001; \\
  A. Morales {\em et al.}, Phys. Lett. B {\bf 489} (2000) 268; \\
  L. Baudis {\em et al.}, Phys. Rev. D {\bf 63} (2001) 022001.
\bibitem{bib-nai}
  R. Bernabei {\em et al.}, Phys. Lett. B {\bf 389} (1996) 757; \\
  P.F. Smith {\em et al.}, Phys. Lett. B {\bf 379} (1996) 299;\\
  G. Gerbier {\em et al.}, Astropart. Phys. {\bf 11} (1999) 287; \\
  K. Fushimi {\em et al.}, Astropart. Phys. {\bf 12} (1999) 185.
\bibitem{bib-dama}
  R. Bernabei {\em et al.}, Phys. Lett. B {\bf 480} (2000) 23.
\bibitem{bib-cdms}
  R. Abusaidi {\em et al.}, Phys. Rev. Lett. {\bf 84} (2000) 5699.
\bibitem{bib-cresst}
  M. Bravin {\em et al.}, Nucl. Instr. Meth. A {\bf 444} (2000) 323.
\bibitem{bib-navick}
  X.F. Navick {\em et al.}, Nucl. Instr. Meth. A {\bf 444} (2000) 361.
\bibitem{bib-distefano}
  P. Di Stefano {\em et al.}, Astropart. Phys. {\bf 14} (2001) 329,
  astro-ph/0004308.
\bibitem{bib-benoit}
  A. Benoit {\em et al.}, Phys. Lett. B {\bf 479} (2000) 8,
  astro-ph/0002462.
\bibitem{bib-chazal}
  V. Chazal {\em et al.}, Astropart. Phys. {\bf 9} (1998) 163.
\bibitem{bib-cryostat}
  A. de Bellefon {\em et al.}, Astropart. Phys. {\bf 6} (1996) 35.
\bibitem{bib-luke}
  M.P. Chapellier {\em et al.}, Physica B {\bf 284-288} (2000) 2135;\\
  P.N. Luke, J. Appl. Phys. {\bf 64} (1988) 6858; \\
  B. Neganov and V. Trofimov, USSR patent No 1037771, 1981;
  Otkrytia i izobreteniya {\bf 146} (1985) 215.
\bibitem{bib-qn}
  L. Baudis {\em et al.}, Nucl. Instr. Meth. A {\bf 418} (1998) 348.
\bibitem{bib-geant}
 R. Brun, F. Bruyant, M. Maire, A.C. McPherson, and P. Zanarini,
 {\it GEANT3}, CERN Report DD/EE/84-1 (1987).
\bibitem{bib-nuclide}
  M.C. Lederer and V.S. Shirley, Table of isotopes
  (VII$^{th}$ edition), John Wiley \& Sons (New York) 1978.
\bibitem{bib-helm}
  R.H. Helm, Phys. Rev. {\bf 104} (1956) 1466;\\
  J. Engel, Phys. Lett. B {\bf 264} (1991) 114.

\end{thebibliography}
\end{document}